\documentclass[11pt,preprintnumbers,aps,amssymb,nofootinbib,amsmath]
{revtex4}
\usepackage{epsfig,epsf}
\usepackage{xfrac} 
\usepackage{bm} 
\newcommand{\beq}{\begin{equation}}
\newcommand{\beql}[1]{\begin{equation}\label{#1}}
\newcommand{\eeq}{\end{equation}}
\newcommand{\eq}[1]{(\ref{#1})}
\newcommand{\fig}[1]{Fig.~\ref{#1}}
\renewcommand{\sec}[1]{Sec.~\ref{#1}}
\newcounter{topiccounter}
\renewcommand{\b}[1]{{\bm #1}} 
\newcommand{\unit}[1]{\hat {{\bm #1}}} 

\newcommand{\as}{\alpha_s}

\newcommand{\e}{\varepsilon}

\newcommand{\jpsi}{J\mskip -2mu/\mskip -0.5mu\psi}

\begin{document}

\title{Quarkonium dissociation  in quark-gluon plasma via ionization in magnetic field}

\author{Kevin Marasinghe and Kirill Tuchin$\,^{a}$\\}

\affiliation{
$^a\,$Department of Physics and Astronomy, Iowa State University, Ames, IA 50011}

\date{\today}


\begin{abstract}

We study the impact of a magnetic field, generated in collisions of relativistic heavy ions, on the decay probability of a quarkonium produced in the central rapidity region.  The quark and antiquark components are subject to mutually orthogonal electric and magnetic fields in the quarkonium comoving frame. In the presence of an electric field quarkonium has a finite dissociation probability. We use the WKB approximation to derive the dissociation probability. We found that quarkonium dissociation energy, i.e. the binding energy at which the dissociation probability is of order unity, increases with the magnetic field strength. It also increases with quarkonium momentum in the laboratory frame due to Lorentz boost of electric field in the comoving frame. We argue that $\jpsi$'s produced in heavy-ion collisions at LHC  with  $P_\bot>9$~GeV would dissociate even in vacuum. In plasma, $\jpsi$ dissociation in a magnetic field is much stronger due to the decrease  of its binding energy with temperature. We discuss phenomenological implications of our results. 

\end{abstract}

\maketitle

\section{Introduction}\label{sec:intr}

In their seminal 1986 paper \cite{Matsui:1986dk} Matsui and Satz suggested that quarkonium production in high energy nucleus--nucleus collisions can be used as an indicator for production of quark-gluon plasma (QGP). They argued that a quarkonium of radius $r$ will dissociate in QGP when the Debye screening radius $r_D(T)$ becomes smaller then $r$. The observed effect would be an ``anomalous" suppression of  quarkonium yield. In practice, determination of the screening radius $r_D(T)$ turned out to be a remarkably difficult problem even in a static medium. The existing  approaches to solve this problem include  lattice QCD calculations of quarkonium correlators \cite{Umeda:2002vr,Asakawa:2003re,Datta:2003ww,Jakovac:2006sf,Aarts:2007pk,Aarts:2010ek}, construction of potential models of quarkonium spectral functions \cite{Digal:2001ue,Wong:2004zr,Mocsy:2005qw,Alberico:2006vw,Cabrera:2006wh,Mocsy:2007yj,Mocsy:2007jz}, and use of effective field theory \cite{Laine:2007qy,Laine:2007gj,Brambilla:2008cx}. It is remarkable that in spite of much progress there still exists  substantial uncertainty in the value of the $\jpsi$ dissociation temperature and in the functional form of $r_D(T)$, see e.g.\ \cite{Satz:2005hx,Zhao:2010nk}.  Another complication arises due to  ``cold nuclear matter effects", i.e.\ nuclear effects that are independent of temperature and that would occur even if no plasma were formed. There is ongoing controversy as to the nature of these effects, although there is  agreement that they lead to suppression of quarkonium yield \cite{Faccioli:2008ir,Vogt:2001ky,Eskola:2009uj,Frawley:2008kk,Kharzeev:2003sk,Kharzeev:2008nw,Kharzeev:2008cv,Tuchin:2009tz,Kharzeev:1995br,Blaizot:1988hh}. 

It has been recently realized \cite{Kharzeev:2007jp} that colliding heavy-ions produce a very strong magnetic field in the direction perpendicular to the reaction plane (defined as a plane containing momenta of the ions and the impact parameter). This has a number of interesting phenomenological consequences \cite{Kharzeev:2007jp,Tuchin:2010vs,Tuchin:2010gx,Mohapatra:2011ku}. The strength of  the magnetic field  at RHIC is estimated as $eB_0\approx m_\pi^2$, and at LHC as $eB_0\approx 15\,m_\pi^2$ \cite{Kharzeev:2007jp,Skokov:2009qp}.  The relaxation time of the magnetic field neglecting the back reaction is of the order of  the width of the two Lorentz contracted nuclei divided by $c$, which is $\sim 0.1$~fm at RHIC and $\sim0.01$~fm at LHC. However, we argued recently \cite{Tuchin:2010vs}  that  the relaxation time of this field is actually much larger if the back reaction is taken into account.
\footnote{
This can be seen as follows. Upon collision, the magnetic field $\b B_0$ rapidly decreases with time inducing an electric field  of strength $ E\sim B_0R/\tau$ circulating around the direction of $\b B_0$; here $R$ is the nuclear radius and $\tau$ the relaxation time. This electric field generates circular currents that, by  Lenz law, support the original magnetic field. There are two types of generated currents: (i)  Foucault currents in the QGP;  (ii) Current of charged fermions produced via the Schwinger mechanism \cite{Schwinger:1951nm}. Note that lepton Schwinger pairs are certainly not in equilibrium with the QGP and therefore do not contribute to the plasma electrical conductivity. In the first case, for a medium with electrical conductivity $\sigma$ we estimate $B_0\sim jR\sim \sigma E R\sim \sigma B_0 R^2/\tau$, which implies 
\beql{tau1}
\tau \sim \sigma R^2\,.
\eeq
Using the values of the electrical conductivities obtained in \cite{Gupta:2003zh} and in \cite{Aarts:2007wj,Ding:2010ga} we get $\tau\sim 8$~fm and $\tau\sim 0.4$~fm correspondingly. 
In the second case, recalling that the density of the Schwinger pairs is $n\sim (eE)^2\tau$ we have
$B_0\sim en vR\sim e^3B_0^2R^3/\tau$, which implies 
\beql{tau2}
\tau \sim e^2(eB_0)R^3\,.
\eeq
We took into account that the Schwinger pairs are relativistic $v\sim 1$ since their longitudinal momentum is of order $eE\tau \sim eB_0R\gg m_e$ for $R=5\,\text{fm}$. We estimate that at RHIC $\tau \sim 0.5$~fm, while at LHC $\tau\sim 8$~fm. A more accurate estimate of $\tau$ requires not only better knowledge of electrical conductivity in expanding plasma, but also numerical solution of the magneto-hydrodynamic equations.} 
In fact, the magnetic field can be treated as static if the distance over which it significantly varies is much larger than the quarkonium radius. We will refer to this  approximation as ``quasi-static". For a   quarkonium with binding energy $\e_b$ and radius $\as/\e_b$,  the quasi-static approximation applies when $\e_b\tau/\as\gg 1$. For $\tau =2$~fm we get $\e_b\tau/\as\approx 23$, which is comfortably large enough to justify the quasi-static approximation, where we assumed that $\e_b$ is given by its vacuum value.  As $T$ increases $\e_b$ drops, while $\tau$ increases. Temperature dependence of $\e_b$ is model dependent, however it is certain that eventually it vanishes at some finite temperature $T_0$. Therefore, only in  the close vicinity of $T_0$, i.e.\ at very small binding energies, the quasi-static approximation is not applicable.  We thus rely on  the quasi-static approximation throughout the paper. 

\bigskip
A magnetic field has a three-fold effect on quarkonium:
\begin{enumerate}
\item \emph{Lorentz ionization}. Suppose quarkonium travels with constant velocity in a magnetic field in the laboratory frame. Boosting to the quarkonium  comoving  frame, we find mutually orthogonal electric and magnetic fields given by Eqs.~\eq{boost},\eq{lt}. In the presence of  an electric field quark and antiquark have a finite probability to tunnel through the potential barrier thereby causing  quarkonium dissociation. 
In atomic physics such  a process is referred to as Lorentz ionization. In the non-relativistic approximation, the tunneling probability is of order unity when the electric field $E$ in the comoving frame satisfies $eE\gtrsim m^{1/2}\e_b^{3/2}$ (for weakly bound states), where $\e_b$ is  binding energy and $m$ is quark mass, see \eq{fnr-gsmall}. This effect causes a significant increase in quarkonium dissociation energy $\e_d$. The corresponding results for $\jpsi$ are exhibited in \fig{fig:dissoc}. They suggest that $\jpsi$ in plasma dissociates at a much \emph{lower} temperature then it would in the absence   of a magnetic field.

\item \emph{Zeeman effect.} Energy of a quarkonium state depends on  spin $S$, orbital angular momentum $L$, and total angular momentum $J$. In a magnetic field these states split; the splitting energy in a weak field is  $\Delta M = \frac{eB_0}{2m}g J_z$, where $J_z=-J,-J+1,\ldots, J$ is  projection of the total angular momentum on the direction of magnetic field, $m$ is quark mass and $g$ is Land\'e factor depending on $J$, $L$ and $S$ in a well-known way, see e.g.\ \cite{LL3-113}.  For example, $\jpsi$ with $S=1$, $L=0$ and $J=1$ ($g\approx 2$) splits into three states with $J_z= \pm 1, 0$ and with mass difference $\Delta M=0.15$~GeV, 
where we used $eB_0=15m_\pi^2$ (the value of magnetic field expected at LHC \cite{Kharzeev:2007jp,Skokov:2009qp}). Thus,  the Zeeman  effect leads to the emergence of new quarkonium states in plasma. 

\item \emph{Distortion of the quarkonium potential} in magnetic field. This effect arises in higher order  perturbation theory and becomes important at field strengths of order $B\sim 3\pi m^2/e^3$  \cite{Machet:2010yg}. This  is $3\pi/\alpha$ times stronger than the critical Schwinger's field. Therefore, this effect can be neglected at the present collider energies. 

\end{enumerate}
 
The purpose of this paper is the analysis of the Lorentz ionization  of quarkonia.

\bigskip
 Our paper is organized as follows. 
 In \sec{sec:ion.rate} we give fully relativistic derivation of the Lorentz ionization probability $w$ using the WKB approximation.  We treat  the force binding $q$ and $\bar q$ into quarkonium as a short-range one i.e.\  $(M \e_b)^{1/2} R \ll1$, where $\e_b$ and  $M$ are  binding energy and mass of  quarkonium, respectively,    and $R$ is the radius of the nuclear force  given by $R\approx (\as/\sigma)^{1/2}$, where $\sigma= 1$~GeV/fm is the string tension. For example, the binding energy of $c$ and $\bar c$ in $\jpsi$ in vacuum is  $\e_b=0.64~\text{GeV}\ll M/R^2= M\sigma/\as \approx 3$~GeV. This approximation is even better at finite temperature on account of $\e_b$ decrease.  Regarding $\jpsi$ as being bound by a short-range force enables us to 
 calculate the dissociation probability $w$ with exponential accuracy $w\approx e^{-f}$, independently of the precise form of the quarkonium wave function.  This is especially important since solutions of the relativistic two-body problem for quarkonium are not readily available.  We will argue that exponential accuracy is adequate for determining the quarkonium ionization energy due to rapid  variation of $f$ with binding energy. Our result for $w$ is in agreement with previous calculations \cite{Popov:1998aw}.

In \sec{sec:nr} we study  non-relativistic approximation of  ionization probability $w$. We argue that  it provides a remarkably good estimate of relativistic formulas, see \fig{fig:phi}. This is an important observation as it allows us to include the contribution  of the quark spin interaction with the magnetic field. A fully relativistic calculation that accounts for the spin contribution is not yet available. This is discussed in \sec{sec:spin}. In \sec{sec:nr}  we also show that relativistic formulas for $w$ derived in \sec{sec:ion.rate} reduce to  non-relativistic formulas found in the literature.  

In \sec{sec:dis.energy} we calculate the dissociation energy of $\jpsi$ in a magnetic field and claim that 
it strongly depends on magnetic field and $\jpsi$'s velocity $V$ in the lab frame. Our results are summarized in \fig{fig:dissoc}. In a strong magnetic field such as the one expected to be produced at the LHC, a $\jpsi$ moving with $P_\bot>9$~GeV in the reaction plane is expected to dissociate because of  magnetic field even in vacuum. At finite temperature, when  binding energy decreases, dissociation becomes prominent at lower transverse momenta. The magnetic field has no influence on $\jpsi$ dissociation when it moves perpendicularly to the reaction plane (i.e.\ parallel to the field) because the corresponding electric field vanishes in the comoving frame.  Non-trivial azimuthal angle dependence of $w$ may be an important  source of azimuthal anisotropy of $\jpsi$ production in heavy ion collisions. 

In \sec{sec:disc} we discuss the phenomenological significance of our results.

\bigskip
Before continuing with the main part of the paper, we list here some of our notational definitions in order to avoid confusion:   $\b V$ and $\b P$ are velocity and momentum of quarkonium in the lab frame;  $M$ is its mass; $\b p$ is the momentum of quark or anti-quark  in the comoving frame;  $m$ is its mass;  $\b B_0$ is the magnetic field in the lab frame, $\b E$ and $\b B$ are  electric and magnetic fields in the comoving frame; $\gamma_L$ is the quarkonium Lorentz factor; and  $\gamma$ is a parameter defined in \eq{gamma}. We use Gauss units throughout the paper; note that expressions $eB$, $eE$ and $eB_0$ are the same in Gauss and Lorentz-Heaviside units.

\section{Quarkonium ionization rate}\label{sec:ion.rate}

\subsection{Comoving frame}\label{sec:com}

Consider a quarkonium traveling with velocity $\b V$ in constant magnetic field $\b B_0$.  Let $\b B $ and $\b E$ be magnetic and electric fields in the comoving frame, and let  subscripts $\parallel$ and $\bot$ denote  field components parallel and perpendicular  to $\b V$ correspondingly. Then,
\begin{subequations}\label{boost}
\begin{align}
&E_\parallel = 0\,,    & \b E_ \bot=\gamma_L \b V\times \b B_0\,,  \\
& B_\parallel = \frac{\b B_0\cdot \b V}{V}\,, & \b B_ \bot =\gamma_L \frac{(\b V\times \b B_0)\times \b V}{V^2}\,,
\end{align}
\end{subequations}
where $\gamma_L=(1-V^2)^{-1/2}$. Clearly, in the comoving frame $\b B\cdot \b E=0$. If quarkonium travels at angle $\phi$ with respect to the magnetic field in the laboratory frame, then
\begin{align}\label{lt}
& B= B_0\sqrt{\cos^2\phi(1-\gamma_L^2) +\gamma_L^2}\,, &E=B_0\gamma_L V\sin\phi\,.
\end{align}
We choose $z$ and $y$ axes of the comoving frame such that $\b B= B\unit z$ and $\b E= E\unit y$. A convenient gauge choice is $\b A= -B y\, \unit x$ and $\varphi=-Ey$. For a future reference we also define a useful dimensionless parameter $\rho$ \cite{Popov:1997-A}
\begin{align}\label{rho}
\rho= \frac{E}{B}= \frac{\gamma_L V\sin\phi}{\sqrt{\cos^2\phi(1-\gamma_L^2) +\gamma_L^2}}\,.
\end{align}
Note, that (i) $0\le \rho\le 1$ because $B^2- E^2 =  B_0^2 \ge 0$ and (ii) when quarkonium moves perpendicularly to the magnetic field $\b B_0$, $\rho=V$. 

\subsection{WKB method}\label{sec:wkb}

It is natural to study quarkonium ionization in the comoving frame. As explained in the Introduction, ionization is quantum tunneling through the potential barrier caused by the electric field $\b E$. In this subsection we employ the quasi-classical, a.k.a.\ WKB, approximation to calculate the quarkonium decay probability $w$.  For the gauge choice specified in \sec{sec:com} quark energy $\e_0$ ($\e_0<m$) in electromagnetic field can be written as
\beql{hamilt}
 \e_0= \sqrt{m^2+(\bm p-e \bm A)^2}+e\varphi=\sqrt{m^2+(p_x+eBy)^2+p_y^2+p_z^2}-eEy\,.
\eeq
In terms of $\e_0$, quarkonium binding energy is  $\e_b=m-\e_0$. To simplify notations, we will set $p_z=0$, because the quark moves constant momentum along the direction of magnetic field.

\begin{figure}[ht]
      \includegraphics[height=5.5cm]{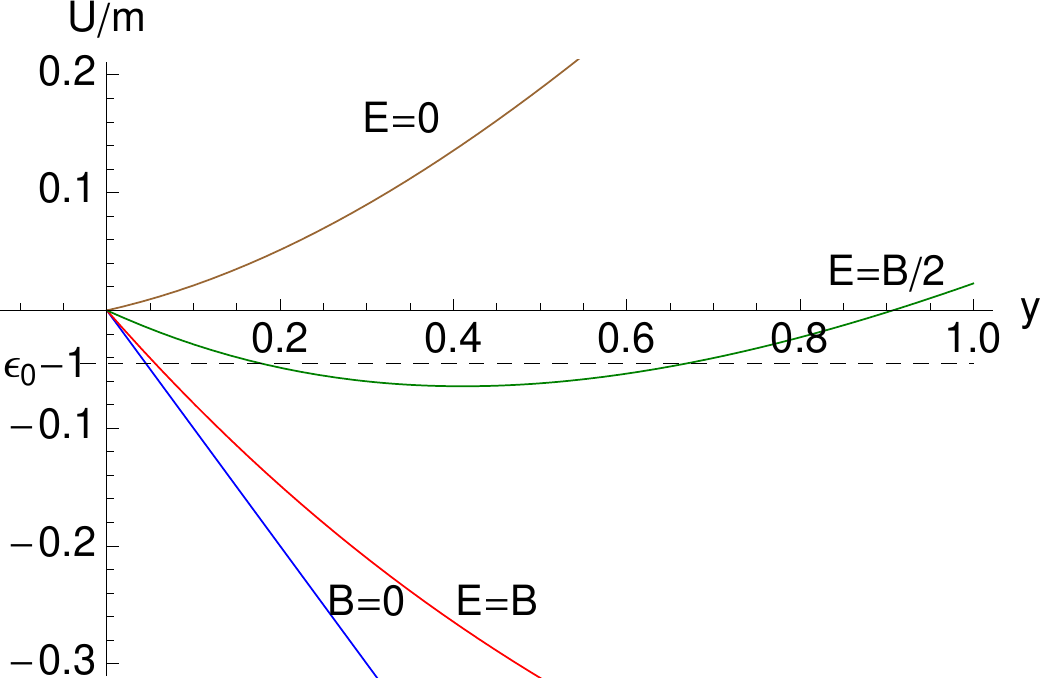} 
        \caption{Effective potential $U(y)=\sqrt{m^2+(p_x+eBy)^2+p_y^2}-eEy-\sqrt{m^2+p_x^2+p_y^2}$ for $p_y=0$, $p_x=m/6$, $B=m$ (except the blue line where $B=0$). The width of the potential barrier decreases with $E$ and increases with $B$. $1-\epsilon_0$ corresponds to the binding energy in units of $m$. }
\label{fig:V}
\end{figure}

The effective potential $U(y)= \e_0(y)-\sqrt{m^2+\b p^2}$ corresponding to  \eq{hamilt} is plotted in \fig{fig:V}.
We can see that the tunneling probability is finite only if $E>0$. It is largest when $B=0$. It has been already noted before in \cite{Popov:1998aw,Popov:1997-A,Popov:1998-A} that the effect of the magnetic field is to stabilize the bound state. In spite of the linearly rising potential (at $B>E$) tunneling probability is finite as the result of rearrangement of the QED vacuum in electric field. 

Ionization probability of quarkonium equals its tunneling probability through  the potential barrier. The later is given by the transmission coefficient 
\beql{w}
w= e^{-2\int_0^{y_1}\sqrt{-p_y^2}dy}\equiv e^{-f}\,.
\eeq
In the non-relativistic approximation one can also calculate the pre-exponential factor, which appears due to the deviation of the quark wave function from the quasi-classical approximation. Such a calculation requires matching quark wave functions inside and outside the barrier \cite{LL3-77}. To determine the pre-exponential factor in the relativistic case one needs to solve the relativistic two-body problem, which is analytically challenging \cite{scott}. Fortunately, as we argue later, one does not need to know the pre-factor to make reliable estimates of the quarkonium dissociation energy.  

We now proceed with the calculation of function $f$. Since $B>E$ Eq.~\eq{hamilt} can be written as 
\beql{py1}
p_y^2=-e^2(B^2-E^2)(y-y_1)(y-y_2)\,,
\eeq
where 
\beql{y12}
y_{1,2}=\frac{\e_0 E-p_x B\mp \sqrt{(\e_0 E-p_xB)^2-(B^2-E^2)(-\e_0^2+m^2+p_x^2)}}{e(B^2-E^2)}\,.
\eeq
Define dimensionless variables $\epsilon_0 = \e_0/m$ and $q=p_x/m$. Integration in \eq{w} gives:
\begin{align}\label{f}
\frac{f}{m^2}&=\frac{\sqrt{-\epsilon_0^2+1+q^2}(\epsilon_0 E-q B)}{e(B^2-E^2)}\nonumber\\
&-\frac{(\epsilon_0 E-q B)^2-(B^2-E^2)(-\epsilon_0^2+1+q^2)}{e(B^2-E^2)^{3/2}}
\ln\left\{
\frac{\epsilon_0 E-q B+\sqrt{(B^2-E^2)(-\epsilon_0^2+1+q^2)}}{\sqrt{(\epsilon_0 E-qB)^2-(B^2-E^2)(\epsilon_0^2+1+q^2)}} 
\right\}\,.
\end{align}
For different $q$'s $w=e^f$ gives the corresponding ionization probabilities. The largest probability corresponds to smallest $f$, which occurs at  momentum $q_m$ determined by equation \cite{Popov:1998aw}
\beql{q-min}
\frac{\partial f(q_m)}{\partial q_m}=0\,.
\eeq
Using \eq{f} and  parameter $\rho$ defined in \eq{rho} we find
\beql{f'}
\frac{\rho(\epsilon_0 -\rho q_m)}{1-\rho^2}\ln\left\{\frac{\epsilon_0\rho-q_m+\sqrt{1-\rho^2}\sqrt{-\epsilon_0^2 +1 +q_m^2}}{\sqrt{(\epsilon_0-\rho q_m)^2-1+\rho^2}}\right\}=\frac{\sqrt{-\epsilon_0^2 +1+q_m^2}}{\sqrt{1-\rho^2}}\,.
\eeq
This is an implicit equation for  the extremal momentum $q_m=q_m(\epsilon_0,\rho)$. Substituting $q_m$ into \eq{f} one obtains $f=f(\epsilon_0,\rho)$, which by means of \eq{w} yields the ionization probability.  The quasi-classical approximation that we employed in this section is valid inasmuch as  $f(q_m)\gg 1$. 

In order to compare with the results obtained in  \cite{Popov:1998aw} using the imaginary time method, we can re-write Eq.~\eq{f'}  in terms of an auxiliary parameter  $\tau_0$ as
\begin{subequations}
\begin{align}
&
\label{tau}
\tau_0=\frac{\sqrt{1-\rho^2}\sqrt{-\epsilon_0^2 +1 +q_m^2}}{\rho(\epsilon_0 -\rho q_m)}\,,\\
&\label{f'2}
\frac{\tanh\tau_0}{\tau_0}=\rho\,\frac{\epsilon_0-\rho q_m}{\epsilon_0 \rho -q_m}\,.
\end{align}
\end{subequations}
Taking advantage of these equations, Eq.~\eq{f} can be cast into a more compact form 
\beql{f2}
f_m=\frac{m^2\tau_0\rho}{eE\sqrt{1-\rho^2}}[1-\epsilon_0(\epsilon_0-q_m\rho)]\,,
\eeq
where we denoted $f_m=f(q_m)$.
 This agrees with results of Ref.~\cite{Popov:1998aw}. Numerical solution of \eq{tau},\eq{f'2},\eq{f2} is shown in \fig{fig:phi}.

\subsection{Special case:  Crossed fields}
An important limiting case is crossed fields $E=B$. Since also  $\b E\perp \b B$, see \sec{sec:com},
both field invariants vanish. Nevertheless, quarkonium ionization probability
is finite \cite{Popov:1998aw}. This limit is obtained by  taking $\rho\to 1$ in the equations from the previous section. Employing \eq{tau} and \eq{f'2} we get the following condition for extremum
\beql{r1-ext}
\epsilon_0^2-1+2q_m^2-3\epsilon_0 q_m=0\,,
\eeq
with the solution
\beql{r1-e-s}
q_m= \frac{1}{4}\left( 3\epsilon_0-\sqrt{\epsilon_0^2+8}\right)\,.
\eeq
Substituting into \eq{f2} produces
\beql{fr-1}
f_m= \frac{2}{3}\frac{m^2}{eE}\frac{(-\epsilon_0^2+1+q_m^2)^{3/2}}{\epsilon_0-q_m}\,.
\eeq

\section{Non-relativistic approximation}\label{sec:nr}

A very useful approximation of the relativistic formulas derived in the previous section is  the non-relativistic limit because (i) it provides a very good numerical estimate, see \fig{fig:phi}, (ii) it allows 
us to eliminate the parametric dependence in \eq{f},\eq{f'} and write $f(q_m)$ explicitly in terms of $\rho$ and $\epsilon_0$, and (iii) spin effects can be accounted for.

\subsection{Arbitrary binding}\label{sec:arb}

Motion of a particle can be treated non-relativistically if its momentum is much less than its mass. In such a case  $\e_0\approx m$ or \ $\e_b=m-\e_0\ll m$.  Additionally, motion of a charged particle in electromagnetic field is non-relativistic if  $E\ll B$. Indeed, the average velocity of a non-relativistic particle is of order $v\sim E/B=\rho$. Thus,  the non-relativistic limit is obtained by taking the limits $\epsilon_b=\e_b/m\ll 1$ and $\rho\ll 1$. In these limits the extremum conditions \eq{tau},\eq{f'2} reduce to 
\begin{subequations}
\begin{align}
&\tau_0= \frac{\sqrt{2\epsilon_b+q_m^2}}{\rho}\,,\label{nr-tau-1}\\
& \frac{\tanh\tau_0}{\tau_0}=\frac{\rho}{\rho-q_m}\label{nr-tau-2}\,.
\end{align}
\end{subequations}
Out of two solution to \eq{nr-tau-1} we pick the following one
\beql{qm}
q_m= -\sqrt{\tau_0^2\rho^2-2\epsilon_b}\,.
\eeq
The sign of $q_m$ is fixed using \eq{nr-tau-2} by noticing that $\tanh \tau_0/\tau_0<1$.  Eliminating $q_m$ gives:
\beql{nr2}
\tau_0^2-(\tau_0\coth\tau_0-1)^2=\gamma^2\,,
\eeq
where 
\beql{gamma}
\gamma = \frac{\sqrt{2\epsilon_b}}{\rho}\,.
\eeq
$\gamma$ is analogous to the  adiabaticity parameter of Keldysh \cite{Keldysh-ioniz}. Taking the non-relativistic limit of \eq{f2} and using \eq{qm} yields
\beql{fnr}
f_m=\frac{2m^2(2\epsilon_b)^{3/2}}{3eE}g(\gamma)\,,
\eeq
where $g(\gamma)$ is  the  Keldysh function \cite{Keldysh-ioniz} 
\beql{keldysh}
g(\gamma)=\frac{3\tau_0}{2\gamma}\left[ 1-\frac{1}{\gamma}\left( \frac{\tau_0^2}{\gamma^2}-1\right)^{1/2}\right]\,.
\eeq

\begin{figure}[ht]
      \includegraphics[height=6cm]{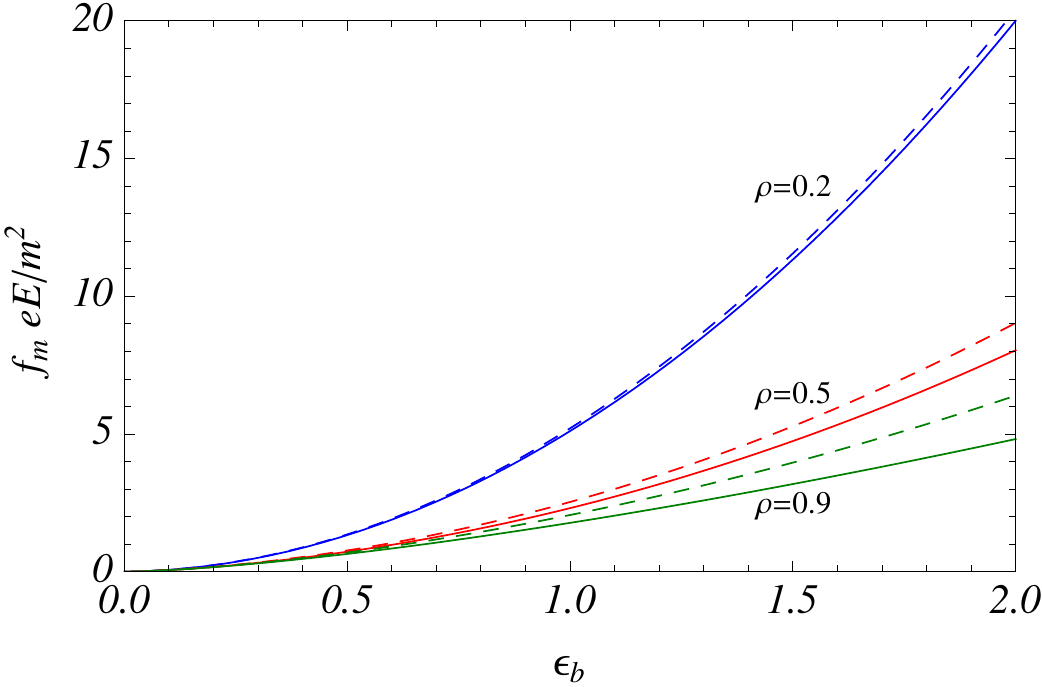} 
        \caption{Dimensionless function $f_m eE/m^2$ versus $\epsilon_b$ for different values of $\rho$. The solid line is the full relativistic calculation, the dashed line is the non-relativistic approximation. $\jpsi$ binding energy in vacuum corresponds to $\epsilon_b=0.68$. }
\label{fig:phi}
\end{figure}

In \fig{fig:phi} we show the dimensionless ratio $f_meE/m^2$ as a function of the binding energy $\epsilon_b$ (in units of $m$) for several values of  $\rho$. The vacuum binding energy of $\jpsi$ corresponds to $\epsilon_b = 0.68$. We observe an excellent agreement between the full relativistic calculation and  the non-relativistic approximation.  At $\rho=0.9$ and $\epsilon_b = 0.68$ the difference between the two lines is 10\% and can be further improved by considering higher order corrections to $f_m$ \cite{Popov:1998-A}.

\subsection{Weak binding}\label{sec:weak}

Of special interest is the limit of  weak binding $\gamma\ll 1$, i.e.\ $\sqrt{2\epsilon_b}\ll \rho$. Expanding \eq{nr2} at small $\gamma$  and $\tau_0$ yields
\beql{nr-gsmall}
\tau_0=\gamma\left(1+\frac{1}{18}\gamma^2\right)\,
\eeq
and substituting into \eq{keldysh} and subsequently into  \eq{fnr} yields
\beql{fnr-gsmall-0}
f_m=\frac{2}{3}\frac{m^2}{eE}(2\epsilon_b)^{3/2}\,.
\eeq
Hence, the quarkonium dissociation probability reads \cite{LL3-77}
\beql{fnr-gsmall}
w=   \exp\left\{-\frac{2}{3}\frac{(2\e_b m)^{3/2}}{meE}\right\}\,.
\eeq
Since the quasi-classical approximation employed in this paper is valid if $f(q_m)\gg 1$, it follows that 
the binding energy must satisfy 
\beql{cond-1}
 \frac{(eE)^{2/3}}{m^{1/3}}\ll \e_b\ll \rho^2m\,.
\eeq
Note also that we work in the approximation of the short-range binding potential meaning that  $\sqrt{2\epsilon_b} \ll 1/(mR)$, see \sec{sec:intr}.

\subsection{Strong binding}\label{sec:strong}

In the limit $\gamma\gg 1$,  \eq{nr2} and \eq{keldysh} imply that 
\beql{str-g}
\tau_0=\frac{\gamma^2}{2}\,,\quad g(\gamma)= \frac{3\gamma}{8}\,.
\eeq
Substituting \eq{str-g} into  \eq{fnr} we derive 
\beql{str-f}
f_m= \frac{\e_b^2}{eE}\frac{B}{E}\,.
\eeq
Thus, quarkonium dissociation  probability in the case of strong binding is 
\beql{w-sb}
w=\exp\left\{-\frac{\e_b^2}{eE}\frac{B}{E} \right\}\,.
\eeq
This formula is valid when
\beql{cond-2}
\rho^2m\,,\,\sqrt{eE\rho}\ll \e_b\ll 1/R
\eeq

\subsection{Contribution of quark spin}\label{sec:spin}

So far we have neglected the contribution of quark spin.  In order to take into account the effect of spin interaction with the external field, we can use  squared Dirac equation for a bi-spinor $\psi$: 
\beql{dirac}
\left[ (\e-e\varphi)^2-(\b p-e\b A)^2-m^2+e\b\Sigma\cdot \b B-i e \b \alpha\cdot \b E\right]\psi=0\,,
\eeq
where
\begin{align}
&
\b \Sigma = \left( 
  \begin{array}{cc}
    \b \sigma & 0 \\ 
    0 & \b\sigma 
  \end{array}
  \right)
  &
  \b \alpha = \left( 
  \begin{array}{cc}
    \b \sigma & 0 \\ 
    0 & -\b\sigma 
  \end{array}
  \right)
  \end{align}
Operators $\Sigma_z$ and $\alpha_y$ do not commute. Therefore, in order to apply the WKB method for calculation of the ionization probability one actually needs to square \eq{dirac}, which leads to a differential equation of the fourth order in derivatives. The problem becomes more tractable in the non-relativistic case and for crossed fields. Spin effects in crossed fields were discussed in \cite{popov-review}. 

With  quark spin taken into account, the non-relativistic version of \eq{hamilt} becomes:
\beql{nr-hamilt}
\frac{1}{2m}\left[ (p_x+eBy)^2+p_y^2\right] -eEy-\frac{\mu}{s}\b s\cdot \b B=-\e_b\,,
\eeq
and hence
\beql{py1-nr}
p_y^2= 2m\left(-\e_b+\frac{\mu}{s}\b s\cdot \b B+e E y\right)-(p_x+eBy)^2\,,
\eeq  
where $\mu$ is the quark magnetic moment and $s$ is the projection of spin in the direction of the magnetic field. For a point quark, $\mu=\mu_B= \frac{e\hbar}{2mc}$. 
The effect of quark spin on quarkonium dissociation probability can be taken into account by replacing $\e_b\to \e_b'=\e_b-\frac{\mu}{s}\b s\cdot \b B$ in formulas for $f_m$. With this replacement, all results of this section apply  to a particle with spin. Note that  effective binding energy $\e_b'$ decreases if spin is parallel to the magnetic field and increases if it is antiparallel. In particular, in the case of weak binding
\beql{weak-spin}
w= \sum_{s=\pm 1/2}  \exp\left\{-\frac{2}{3}\frac{(2\e_b m+2seB)^{3/2}}{meE}\right\}\,.
\eeq
Since  the non-relativistic limit provide a good approximation of the full relativistic formulas, we will implement the quark spin dependence using  the non-relativistic prescription.

\section{Dissociation energy of $\jpsi$}\label{sec:dis.energy}

Now, we would like to use the results of the preceding sections to analyze the dissociation probability $w$ of  quarkonium as a function of its binding energy $\e_b$. We suppose that  quarkonium moves with velocity $\b V$ at angle $\phi$ with respect to the external magnetic field $\b B_0$.  Different binding energies may correspond to either different quarkonium states or to the same quarkonium state at different temperatures. 

\begin{figure}[ht]
      \includegraphics[height=7cm]{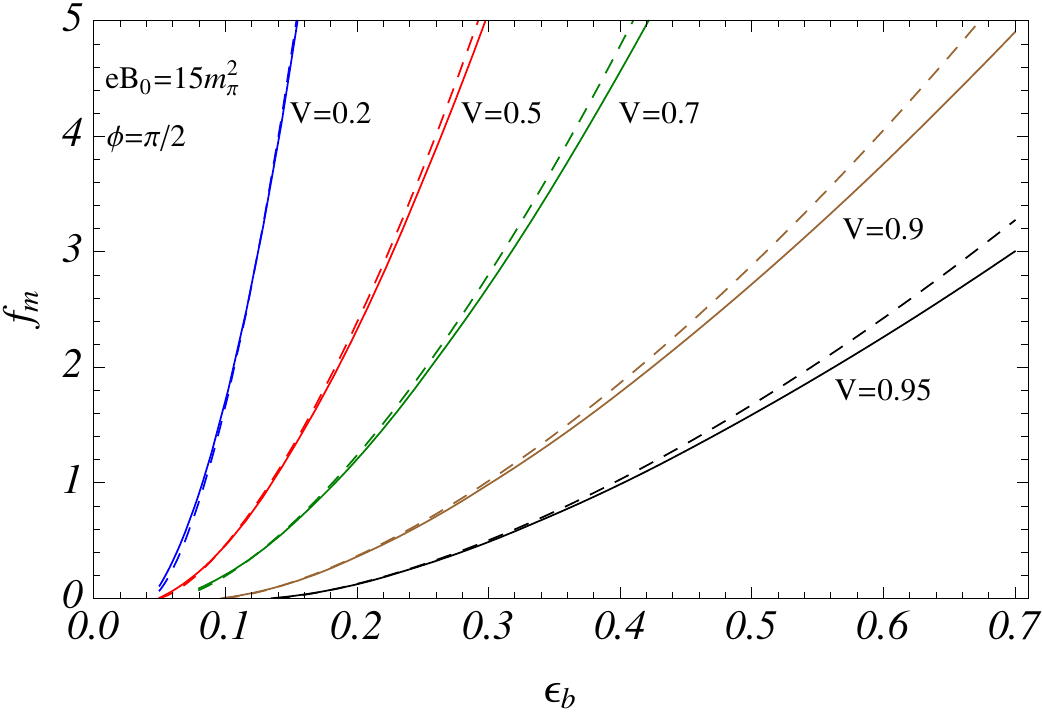} 
        \caption{$f_m $ versus $\epsilon_b$ for different values of $\jpsi$ velocity $V$. Dissociation probability is $w= \exp\{ -f_m\}$. Magnetic field in the lab frame is taken to be $eB_0=15m_\pi^2$. $\jpsi$ moves perpendicularly to the field (i.e.\ in the reaction plane).  Solid lines correspond to the full relativistic calculation, dashed lines to the non-relativistic approximation. $\jpsi$ binding energy in vacuum corresponds to $\epsilon_b=0.68$. }
\label{fig:f-jpsi}
\end{figure}

Dissociation probability is given by \eq{w} as $w=\exp\{-f_m(\epsilon_b)\}$. At $f_m(\epsilon_b)\gg 1$, quarkonium is stable with respect to Lorentz ionization because  $w\ll 1$. The dissociation become probable only at such  values of binding energy $\epsilon_b=\epsilon_d$ that $f_m(\epsilon_d)\sim 1$.  In view of the steep dependence of  $f_m(\epsilon_b)$ on $\epsilon_b$, the probability $w(\epsilon_b)$ is in fact  close to the step function: it turns one when $\epsilon_b< \epsilon_d$ and is zero otherwise. Consequently, we introduce the quarkonium dissociation energy in magnetic field $\e_d=m\epsilon_d$ as the solution to the equation 
\begin{align}\label{ed}
f\left(\epsilon_d-\frac{eB}{2m^2}\right)=1\,,
\end{align} 
where we took into account  the contribution of quark spin, see \sec{sec:spin}. The negative sign in front of the spin term in \eq{ed} is chosen to give the highest dissociation energy. (Which of two particles $c$ or $\bar c$ has higher probability to tunnel depends on the direction of the magnetic field). 
The solution of \eq{ed} depends on electric and magnetic fields in the comoving frame, which in turn depend on the magnetic field in the lab frame $B_0$ and quarkinium kinematics through the Lorentz transformation formulas \eq{lt}. In \fig{fig:f-jpsi} we show function $f_m(\epsilon_b)$ for different  velocities  $V$ of $\jpsi$, assuming it  moves perpendicularly to the magnetic field. In vacuum, $M=3.1$~GeV, $m=1.87$~GeV, so that   $\e_{b}=2m-M=0.64$~GeV. The corresponding dimensionless parameter $\epsilon_b$ varies in the interval $0\le \epsilon_b \le 0.68$.

\begin{figure}[ht]
      \includegraphics[height=7cm]{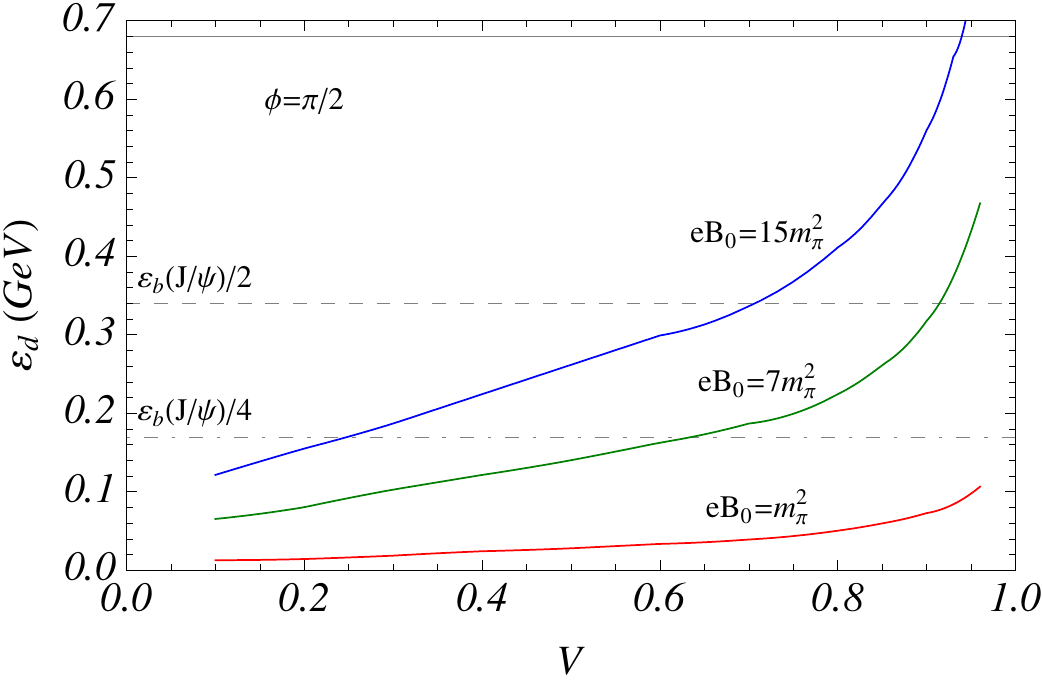} 
        \caption{Dissociation energy of $\jpsi$ as a function of its velocity $V$ in different magnetic fields $B_0$. We assumed that $\jpsi$ moves in the reaction plane. Vacuum dissociation energy is zero. 
The solid horizontal line indicates the vacuum binding energy. Dashed horizontal lines correspond to binding energy at certain temperatures.}
\label{fig:dissoc}
\end{figure}

The numerical solution to Eq.~\eq{ed} is plotted in \fig{fig:dissoc}. We see that the dissociation energy of  $\jpsi$ increases with its velocity and is larger in a strong magnetic field. At $eB_0=15m_\pi^2$, which according to estimates may be reached in heavy-ion collisions at the LHC \cite{Kharzeev:2007jp,Skokov:2009qp}, the dissociation energy equals the vacuum binding energy at $V=0.94$. This implies that most of  $\jpsi$'s moving in the reaction plane with $P_\bot>9$~GeV will dissociate in magnetic field even in vacuum.  

In a thermal medium,  $\e_b$ exhibits strong dependence on temperature $T$. At $T=0$, $\e_b$ equals its vacuum value. As $T$ increases $\e_b$ drops until it vanishes at some $T_0$ near $T_c$. The precise value of $T_0$ as well as the functional form of $\e_b(T)$ is model dependent, see e.g.\ \cite{Satz:2005hx,Zhao:2010nk}.  To illustrate the temperature effect, we indicated in \fig{fig:dissoc} half of the $\jpsi$ binding energy $\e_b/2$ and its quarter $\e_b/4$. Suppose that plasma temperature is such that the binding energy is half of that in vacuum. Then,  at $eB_0=15 m_\pi^2$, all $\jpsi$'s with $V>0.71 $ ($P_\bot>3.1$~GeV) will dissociate, while at $eB_0=7m_\pi^2$ this occurs at $V>0.92$ ($P_\bot>7.3$~GeV).

So far in this section we have discussed the case of quarkonium moving in the reaction plane, i.e.\ at $\phi=\pi/2$. 
Since electric field in the comoving frame is proportional to $\sin\phi$, see \eq{lt}, the dissociation probability decreases at oblique angles. $\jpsi$'s moving parallel to the magnetic field $\b B_0$, i.e.\ at  $\phi=0$, are not affected by electromagnetic field at all. This non-trivial azimuthal angle dependence translates into azimuthal anisotropy of produced $\jpsi$'s, which we plan to address in a separate paper.

The size of the spin contribution is determined by the dimensionless ratio $\xi = eB/2\epsilon_b m^2$, see \eq{weak-spin}. Its magnitude for $\jpsi$ moving with a non-relativistic velocity  perpendicularly to the magnetic field $\b B_0$ is   $\xi\approx (eB_0/240 m_\pi^2)(0.68/\epsilon_b)$. Therefore, even in the strongest field expected at the LHC, $eB_0=15m_\pi^2$, the effect of spin is quite small at $T=0$. $\xi$ increases as the temperature of plasma and/or velocity of $\jpsi$ increases owing to decrease of the binding energy from its vacuum value $\epsilon_b=0.68$ down to zero. When $\xi\sim 1$ non-relativistic treatment of spin breaks down.

\section{Discussion and Conclusions}\label{sec:disc}

In this paper we chose the quarkonium proper frame to calculate the dissociation probability. In this frame 
the potential energy of, say, antiquark (with $e<0$) is a sum of its potential energy in the binding potential and its energy in the electric field $-eEy$, where $y$ is the electric field direction.  Since $|e|Ey$ becomes large and negative at large and negative $y$ (far away from the bound state)  and because the quarkonium potential has finite radius, this region opens up for the motion of the antiquark. Thus  there is a quantum mechanical probability to tunnel through the potential barrier formed on one side by the vanishing quarkonium potential and on the other by increasing absolute value of the antiquark energy in electric field. Of course the total energy of the antiquark (not counting its mass) is negative after tunneling. However, its kinetic energy grows proportionally to $eEy$ as it goes away. By picking up a light quark out of vacuum it can hadronize into a $D$-meson. 
 
If we now go to the reference frame where $E=0$ and there is only magnetic field $B$ (we can always do so since $E<B$), then the entire process looks quite different. An energetic quarkonium travels in external magnetic field and decays into quark-antiquark pair that can late hadronize into $D$-mesons. This happens in spite of the fact that $\jpsi$ mass is smaller than masses of two $D$-mesons due to additional momentum $e\b A$ supplied by the magnetic field. Similarly a photon can decay into electron-positron pair in external magnetic field. 

In the previous section, we observed that $\jpsi$ dissociation energy increases  with magnetic field strength and quarkonium momentum. As a consequence, \emph{quarkonia dissociate at lower temperature} than one would have expected based on calculations neglecting magnetic field. \fig{fig:dissoc} indicates that in heavy-ion collisions at the LHC, $\jpsi$'s moving with $P_\bot > 9$~GeV  in the reaction plane  dissociate with probability of order unity even
 \emph{in vacuum}, where we rely on the magnetic field estimates of \cite{Kharzeev:2007jp}.  This effect is more pronounced at finite temperature owing to the decrease in binding energy and, as seen in \fig{fig:dissoc}, it occurs in weaker fields. 
 
 Although magnetic fields in $pp$ and $pA$ collisions are much weaker than in $AA$ collisions, they are still strong enough to cause $\jpsi$ dissociation at sufficiently high momenta $P_\bot$. A truly spectacular feature of such process would be $\jpsi$ decay into two heavier $D$-mesons. 
 Validity of the quasi-static approximation can perhaps be advocated for a pre-hadronic $c\bar c$ state owing to its short production time. 
This intriguing possibility will be addressed elsewhere. 

The effect of $\jpsi$  dissociation in a magnetic field  vanishes in the direction parallel to the magnetic field, i.e.\ perpendicular to the reaction plane. Therefore,  $\jpsi$ dissociation gives negative contribution to the total azimuthal asymmetry coefficient $v_2$. The size of this contribution is a subject of our ongoing investigation. 

Finally, we would like to mention a possibility of $\jpsi$ dissociation directly by electric fields induced in plasma by the Chiral Magnetic Effect \cite{Kharzeev:2007jp}.\footnote{This observation is due to D. Kharzeev.} We plan to address this interesting possibility in a forthcoming publication.

\acknowledgments
We acknowledge many informative discussions with Dima Kharzeev and useful correspondence with Gert Aarts.  
This work  was supported in part by the U.S. Department of Energy under Grant No.\ DE-FG02-87ER40371.



\end{document}